\documentclass[twocolumn,superscriptaddress,showpacs,prl]{revtex4}

\usepackage[dvips]{graphicx,color}
\usepackage{delarray}
\usepackage{amsmath}
\usepackage{bm}

\newcommand{\ket}[1]{| { #1} \rangle}
\newcommand{\bra}[1]{ \langle {#1}  |}

\begin{document}
\title{Does ``quantum nonlocality without entanglement'' have quantum origin?}

\author{Masato Koashi}
\affiliation{Photon Science Center, University of Tokyo, 2-11-16, Yayoi, Bunkyo-ku, Tokyo 113-8656, Japan}

\author{Koji Azuma}
\affiliation{NTT Basic Research Laboratories, NTT Corporation, 3-1 Morinosato Wakamiya, Atsugi, Kanagawa 243-0198, Japan}

\author{Shinya Nakamura}
\affiliation{Department of Materials Engineering
Science,
Graduate School of Engineering Science, 
Osaka University, 1-3 Machikaneyama, Toyonaka, Osaka 560-8531, Japan}

\author{Nobuyuki Imoto}
\affiliation{Department of Materials Engineering
Science,
Graduate School of Engineering Science, 
Osaka University, 1-3 Machikaneyama, Toyonaka, Osaka 560-8531, Japan}

\
\date{\today}

\begin{abstract}
Quantum separable operations are defined as those
 that cannot produce entanglement from separable states,
and it is known that they strictly surpass  
local operations and classical communication (LOCC) in a number of 
tasks, which is sometimes referred to as ``quantum nonlocality without entanglement.''
Here we consider a task with such a gap 
regarding the trade-off between 
state discrimination and preservation of entanglement.
We show that this task along with the gap 
has an analogue in a purely classical setup,
indicating that the quantum properties are not essential in the
 existence of a nonzero gap between the separable operations and LOCC.
\end{abstract}
\maketitle


The modern and standard definition of entanglement as genuinely quantum correlations
is based on the premise that local operations and 
classical communication (LOCC) are the general 
classical means by which correlations are established.
Entanglement is thus defined to be any correlation that cannot be generated 
from scratch under LOCC.
Reversing this argument, we may also {\em define} a family of general 
classical means, based on the premise that the standard 
definition of entanglement refers to the genuinely quantum correlations,
as the operations that cannot generate entanglement from 
scratch. This family is called separable operations, and has 
a simple mathematical characterization via operators of the namesake 
form. 
Curiously, it has been discovered that not all the separable operations
belong to LOCC, which was first proved in the paper \cite{QNWE} titled
``quantum nonlocality without entanglement.'' The authors discussed
 a task of discriminating nine orthogonal states in a pair of 
three-level systems shared by Alice and Bob. All the states are 
product states and hence perfectly distinguishable by a separable
operation. But they showed that discrimination beyond 
a certain accuracy is never achieved under LOCC regardless of the
number of communication rounds. This limitation
 will be ascribed \cite{QNWE,GV01} to the fact that 
each of Alice and Bob is locally required to distinguish nonorthogonal 
states and hence backaction inevitably causes disturbances.
Similar gaps are later found in discrimination tasks \cite{KTYI07,D09} and in tasks of augmenting a preshared entanglement \cite{CD09,CCL11,CCL12}.

The aim of this paper is to assert that the origin of the gap between 
the separable operations and LOCC is never fully ascribed to 
exclusively quantum properties such as nonorthogonality, measurement
backaction, and entanglement, contrary to what one may infer
from the previous examples. This will be done by raising an example 
of a quantum task and its purely classical analogue, and showing 
that the two tasks share essentially the same gap.
The analogy is established by replacing LOCC to public communication 
(PC) in the presence of a third party. Then, the private correlations
are those that cannot be generated from scratch under PC,
and the classical separable operations are those that 
can never produce private correlations from scratch. 
We will see that the gaps appearing in the quantum and classical tasks
stem from essentially the same origin.

We begin by introducing a quantum task 
for two separated parties, Alice and Bob.
We will proceed to prove the 
existence of a gap before discussing its classical analogue.
Let $\ket{\Phi^\pm}_{AB}$ be Bell states defined by
$\ket{\Phi^\pm}_{AB}:= (\ket{00}_{AB}\pm \ket{11}_{AB})/\sqrt{2}$. 
The goal of Alice and Bob in the following protocol is to 
unambiguously distinguish states $\{\ket{01}_{AB},\ket{10}_{AB}\}$
from $\ket{\Phi^\pm}_{AB}$ with a specified efficiency $Q$
$(0<Q<1)$, 
while minimizing the damage on the entanglement: 
(i) An arbitrator Claire prepares a pair of qubits $AB$ in an initial
 state $\ket{\psi_{\rm ini}}_{AB}$ randomly chosen from 
four candidates
 $\{\ket{01}_{AB},\ket{10}_{AB},\ket{\Phi^+}_{AB},\ket{\Phi^-}_{AB} \}$,
 and sends the qubit $A$ to Alice and $B$ to Bob.
(ii) Alice and Bob perform an operation represented by a set of
Kraus operators $\{ \hat{M}_{\bm{k}}\}_{\bm{k}}$.
If the outcome $\bm{k}$ ensures that $\ket{\psi_{\rm ini}}_{AB}$ was
either $\ket{01}_{AB}$ or $\ket{10}_{AB}$, they so declare and the
protocol ends. This happens when $\bm{k}$ satisfies
$p_{\bm k|\pm}=0$, where 
$p_{\bm k|\pm} := \|\hat{M}_{\bm{k}} \ket{\Phi^\pm}_{AB}\|^2$.
The discrimination must be efficient enough to satisfy 
\begin{equation}
\sum_{\bm{k}:p_{\bm{k}}=0} q_{\bm{k}} \ge Q,
\label{eq:cond-q1}
\end{equation}
where
$q_{\bm{k}}:=(\|\hat{M}_{\bm{k}} \ket{01}_{AB}\|^2
+ \|\hat{M}_{\bm{k}} \ket{10}_{AB}\|^2)/2$
and 
$p_{\bm{k}}:=(p_{\bm{k}|+}+ p_{\bm{k}|-})/2$.
(iii) Otherwise, Claire reveals the identity of $\ket{\psi_{\rm
ini}}_{AB}$. When it was $\ket{\Phi^\pm}_{AB}$, Alice and Bob 
are left with an entangled state 
$\ket{\psi_{\bm k|\pm}}_{AB}:= \hat{M}_{\bm{k}}
\ket{\Phi^\pm}_{AB}/\sqrt{p_{\bm k|\pm}}$. 
The mean residual entanglement is evaluated via the 
following quantity
\begin{equation}
\bar{E}:= \frac{1}{2} \sum_{\bm{k}} \left[p_{\bm{k}|+} {E}(\ket{\psi_{\bm{k}|+}}_{AB}) 
+ 
p_{\bm{k}|-} {E}(\ket{\psi_{\bm{k}|-}}_{AB})\right] , \label{eq:Eq}
\end{equation}
where ${E}(\ket{\psi}_{AB})$ is a measure of entanglement in state
$\ket{\psi}_{AB}$.
${E}(\ket{\psi}_{AB})$ is in general a nondecreasing
function of the smaller Schmidt coefficient $\sqrt{\lambda}$
when $\ket{\psi}_{AB}$ is locally equivalent to $\sqrt{\lambda}\ket{00}_{AB}
+\sqrt{1-\lambda}\ket{11}_{AB}$. If we introduce the concurrence $C:= 2\sqrt{\lambda(1-\lambda)}$ \cite{W98}, we can also regard ${E}$ as
a nondecreasing function ${\cal E}(C)$ of concurrence $C$, i.e., $E={\cal E}(C)$.

Let $\bar{E}_{\rm opt}^{\rm sep}$ be 
the maximum of $\bar{E}$
when the operations in step (ii) are restricted to 
separable operations in the form of 
$\hat{M}_{\bm{k}}=\hat{A}_{\bm{k}} \otimes \hat{B}_{\bm{k}}$.
Let $\bar{E}_{\rm opt}^{\rm LOCC}$ be the supremum over  
the values of $\bar{E}$ achievable via LOCC.
In what follows, we show the existence of a gap between 
$\bar{E}_{\rm opt}^{\rm sep}$ and $\bar{E}_{\rm opt}^{\rm LOCC}$.
This proof is valid for any measure with function ${\cal E}(C)$ 
being strictly concave.
In fact, we will only assume a 
condition weaker than that, i.e., we only require 
such a property at a single point $C=1-Q$, namely,
that there exists $\mu>0$ such that
\begin{equation}
 {\cal E}(C)-\mu(C-1+Q) < {\cal E}(1-Q) \;\; \text{for} \;  C\neq 1-Q,
\label{eq:Econdition2}
\end{equation}
in addition to the continuity and the monotonicity of ${\cal E}(C)$.

An example of an operationally defined measure satisfying 
condition (\ref{eq:Econdition2})
is the maximum success probability ${E}_Q(\ket{\psi}_{AB})$
of converting state $\ket{\psi}_{AB}$ under LOCC to an entangled state 
$\sqrt{\lambda_Q} \ket{00}_{AB} +\sqrt{1-\lambda_Q} \ket{11}_{AB}$
with its concurrence $2\sqrt{\lambda_Q(1-\lambda_Q)}=1-Q$ \cite{V00}.
Since ${E}_Q(\ket{\psi}_{AB})=\min\{1, \lambda_\psi/\lambda_Q\}$ \cite{N99,V99,JP99} with 
$\sqrt{\lambda_\psi}$ being the smaller Schmidt coefficient of $\ket{\psi}_{AB}$,
it is written as a function of the concurrence $C$ of $\ket{\psi}_{AB}$ as
\begin{equation}
 {\cal E}_Q(C)=\frac{1-\sqrt{1-C^2}}{1-\sqrt{1-(1-Q)^2}}
\;\; \text{for} \;  C\le 1-Q
\end{equation}
and ${\cal E}_Q(C)=1$ for $C\ge 1-Q$. It is obvious that ${\cal E}={\cal E}_Q$ satisfies 
Eq.~(\ref{eq:Econdition2}) for 
$0<\mu<1/(1-Q)$.

To prove the existence of a gap,
we first introduce 
$\hat{G}_{\bm{k}}:= \hat{M}_{\bm{k}}^\dag \hat{M}_{\bm{k}}
=\hat{A}_{\bm{k}}^\dag \hat{A}_{\bm{k}}
 \otimes \hat{B}_{\bm{k}}^\dag \hat{B}_{\bm{k}}$,
the element of the positive operator-valued measure (POVM)
for outcome $\bm{k}$.  In the 
matrix representation in the basis 
$\{\ket{i}_A\otimes\ket{j}_B\}_{i,j=0,1}$,
it can be parametrized as 
\begin{align}
\hat{G}_{\bm{k}}= w_{\bm{k}} 
\begin{pmatrix}
1+ x_{\bm{k}} & \xi_{\bm{k}} \\
\xi_{\bm{k}}^* & 1-x_{\bm{k}}
\end{pmatrix}
\otimes 
\begin{pmatrix}
1+ y_{\bm{k}} & \eta_{\bm{k}} \\
\eta_{\bm{k}}^* & 1-y_{\bm{k}}
\end{pmatrix},
\label{eq:matrix}
\end{align}
with $0\le w_{\bm{k}} \le 1$, $-1\le x_{\bm{k}} \le 1$, $-1 \le y_{\bm{k}} \le 1$,
$|\xi_{\bm{k}}|^2\le 1-x_{\bm{k}}^2$, and $|\eta_{\bm{k}}|^2\le 1-y_{\bm{k}}^2$.
Then, the probabilities $p_{\bm{k}}$ and $q_{\bm{k}}$ are written as
\begin{eqnarray}
 p_{\bm{k}}&=& w_{\bm{k}} (1+x_{\bm{k}}y_{\bm{k}}) = p(\hat{G}_{\bm{k}}),
\label{eq:form-of-p}
\\
 q_{\bm{k}}&=& w_{\bm{k}} (1-x_{\bm{k}}y_{\bm{k}}) = q(\hat{G}_{\bm{k}})
\label{eq:form-of-q}
, 
\end{eqnarray}
where $p(\hat{G})$ and $q(\hat{G})$ are functionals that depend linearly on $\hat{G}$ as
$p(\hat{G}):=(\bra{00}\hat{G} \ket{00}+\bra{11}\hat{G} \ket{11})/2$
and $q(\hat{G}):=(\bra{01}\hat{G} \ket{01}+\bra{10}\hat{G} \ket{10})/2$.
The concurrence $C_{\bm{k}|\pm}$ of the final state $\ket{\psi_{\bm{k}|\pm}}_{AB}$ 
is given by 
$C_{\bm{k}|\pm}=|{\rm det}(\hat{A}_{\bm{k}}){\rm det}(\hat{B}_{\bm{k}})|/p_{\bm{k}|\pm} = [{\rm det} (\hat{G}_{\bm k})]^{1/4}/p_{\bm{k}|\pm} $ \cite{W98,V01},
which depends solely on $\hat{G}_{\bm{k}}$.

For the separable operations, we can explicitly give an example 
achieving $\bar{E}={\cal E}(1-Q)$. It has four outcomes $\bm{k}=1,2,3,4$,
and all elements $\{\hat{G}_{\bm{k}}\}$ are {\it diagonal} operators ($\xi_{\bm{k}}=\eta_{\bm{k}}=0$)
specified by 
\begin{eqnarray}
&& w_1=w_2=Q/4, \; w_3=w_4=(2-Q)/4, \nonumber \\
&& x_1=-y_1=-x_2=y_2=1 ,
\label{eq:separable-opt}
\\
&& x_3=y_3=-x_4=-y_4=\sqrt{Q/(2-Q)}.
\nonumber
\end{eqnarray}
Then $q_1=q_2=Q/2$ and $p_1=p_2=0$ satisfy Eq.~(\ref{eq:cond-q1}), 
while
$p_{3|\pm}=p_{4|\pm}=1/2$ and
$C_{3|\pm}=C_{4|\pm}=1-Q$
lead to $\bar{E}={\cal E}(1-Q)$.
Hence,
$\bar{E}_{\rm opt}^{\rm sep}\ge {\cal E}(1-Q)$.

To show the existence of a gap, we need upper bounds on 
$\bar{E}$. First we derive bounds applicable to 
all the separable operations.
The concurrence $C_{\bm{k}|\pm}$ can be bounded via the {\it diagonal} elements of $\hat{G}_{\bm k}$ as 
\begin{equation}
 C_{\bm{k}|\pm}\le C(x_{\bm{k}},y_{\bm{k}}):= \frac{\sqrt{(1-x_{\bm{k}}^2)(1-y_{\bm{k}}^2)}}
{1+x_{\bm{k}}y_{\bm{k}}}, \label{eq:c(x,y)}
\end{equation}
which is easily confirmed by writing down 
$C_{\bm{k}|\pm}/C(x_{\bm{k}},y_{\bm{k}})$ explicitly and 
repeatedly using the inequality
$2\sqrt{|ab|}\le |a|+|b|$. Then we have 
\begin{equation}
\bar{E}\le   
\sum_{\bm{k}:p_{\bm{k}}\neq 0} p_{\bm{k}} {\cal E}(C(x_{\bm{k}},y_{\bm{k}})).
 \label{eq:E}
\end{equation} 
Rewriting the righthand side by using 
\begin{equation}
 \Delta (x,y):= {\cal E}(1-Q)-{\cal E}(C(x,y))-\mu\left( 1-Q-\frac{1-xy}{1+xy}\right),
\end{equation}
we have, after using Eqs.~(\ref{eq:cond-q1}), 
(\ref{eq:form-of-p}), and (\ref{eq:form-of-q}), 
\begin{equation}
 \bar{E}\le  {\cal E}(1-Q)-\sum_{\bm{k}:p_{\bm{k}}\neq 0} p_{\bm{k}} \Delta
  (x_{\bm{k}},y_{\bm{k}}).
\label{eq:difference}
\end{equation}
Rewriting $\Delta(x,y)$, we see that Eq.~(\ref{eq:Econdition2}) implies
\begin{align}
 &\Delta (x,y)\ge 
\mu[(1-xy)/(1+xy)] - \mu C(x,y)   \nonumber \\ 
&=\frac{\mu}{1+xy} \left[(1-xy)-\sqrt{(1-xy)^2-(x-y)^2}\right]
\ge 0,
\end{align}
where both equalities hold only when $C(x,y)=1-Q$ and $x=y$,
namely, $x=y=\pm \sqrt{Q/(2-Q)}$.
Incidentally, this inequality shows that the previous example specified by Eq.~(\ref{eq:separable-opt}) 
is optimal, namely, $\bar{E}_{\rm opt}^{\rm sep}={\cal E} (1-Q)$.

\begin{figure}[t] 
\includegraphics[keepaspectratio=true,height=57mm]{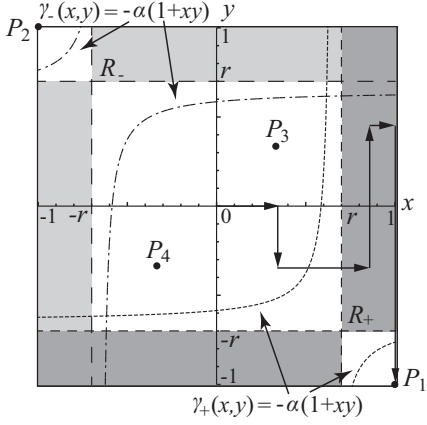}
  \caption{
Trajectory for an outcome $\bm{k}$ of an LOCC operation and
regions $R_{\pm}$. The points $P_3$ and $P_4$
($x=y=\pm \sqrt{Q/(2-Q)}$) are the only points with $\Delta(x,y)=0$,
while only $P_1$ and $P_2$ lead to the success of the discrimination.
We also depict the solutions of $\gamma_\pm(x,y) =- \alpha (1+xy)$.
As an example, we chose the parameters as $Q=0.2$, $r=0.7$, and $\alpha=0.08$.
\label{fig:zigzag}}
\end{figure}

Now let us focus on the properties specific to the LOCC operations.
In an LOCC operation, Alice and Bob alternately send the outcomes 
$k_1,k_2,\ldots,k_n$. The final outcome $\bm{k}$ is regarded as 
the whole of these $n$ outcomes. Let $\hat{G}_{k_1\ldots k_m}$
be the POVM element at the end of the $m$-th round,
and define $x_{k_1\ldots k_m}$ and $y_{k_1\ldots k_m}$ as in 
Eq.~(\ref{eq:matrix}).
Since the $(m+1)$-th measurement does not change $k_1\ldots k_m$,
we have 
$\sum_{k_{m+1}} \hat{G}_{k_1\ldots k_{m+1}} =\hat{G}_{k_1\ldots k_m}$,
implying that the LOCC is a {\em branching} process of the POVM elements.
Consider the branch leading to a final outcome $\bm{k}=k_1k_2\cdots
k_n$, $\hat{1} \to \hat{G}_{k_1} \to \hat{G}_{k_1 k_2} \to
\cdots \to \hat{G}_{k_1\ldots k_n}$,
and the corresponding trajectory of the points on a $xy$-plane,
$(0,0) \to (x_{k_1},y_{k_1}) \to (x_{k_1 k_2},y_{k_1 k_2}) \to
\cdots \to (x_{k_1\ldots k_n},y_{k_1\ldots k_n})$.
Since Alice's measurement does not change the matrix representation 
of Bob's part in Eq.~(\ref{eq:matrix}), the value of $y$ 
keeps on her round. Similarly, Bob's round does not change $x$.
As a result, the trajectory is a zigzag line as in 
Fig.~\ref{fig:zigzag}, 
reflecting the fact that Alice and Bob can refine the POVM 
{\it only alternately}.
These properties \cite{KTYI07} are essential in our proof for the gap.

In what follows, let $\bm{k}=k_1\ldots k_n$ be the final outcome strings with $\hat{G}_{\bm{k}}\neq 0$.
Let us define regions $R_\pm:= \{(x,y)|\gamma_\pm(x,y)\ge 0\}$ 
with
 $\gamma_\pm(x,y):= (x\mp r)(y\pm r)$.
We divide all the final outcome strings $\bm{k}$ 
into three sets $\Gamma_0,\Gamma_+,\Gamma_-$ according to 
the trajectory from $(0,0)$ to $(x_{\bm{k}},y_{\bm{k}})$
by the following rules. 
(i) $\bm{k} \in \Gamma_0$ if the trajectory never goes into the region $R_+\cup R_-$.
(ii) $\bm{k} \in \Gamma_+$ if the first entry point $\hat{G}_{k_1\ldots k_l}$
  into the region $R_+ \cup R_-$ is in the region $R_+$.
(iii) $\bm{k} \in \Gamma_-$ if 
the first entry point $\hat{G}_{k_1\ldots k_l}$
  into the region $R_+ \cup R_-$ is in the region $R_-\backslash R_+$.
Let $\Gamma'_+$ be the set of all intermediate outcome strings $k_1\ldots k_l$
appearing in (ii), and define $\Gamma'_-$ as that for (iii).
Then we have 
\begin{equation}
 \sum_{\bm{k}'\in \Gamma'_\pm}\hat{G}_{\bm{k}'}
= \sum_{\bm{k}\in \Gamma_\pm}\hat{G}_{\bm{k}}. \label{eq:equivalence}
\end{equation}
Since the trajectory $(x,y)$ is a zigzag line, 
any trajectory leading to either $(1,-1)$ or 
$(-1,1)$ must land on 
region $R_+\cup R_-$ (cf. Fig.~\ref{fig:zigzag}).
This implies that $\bm{k}\in \Gamma_++\Gamma_-$
whenever $p_{\bm{k}}=0$. Then, from Eq.~(\ref{eq:cond-q1}), we have
\begin{equation}
 q_++q_- \ge Q \label{eq:eta}
\end{equation}
with $q_\pm := \sum_{\bm{k}\in \Gamma_\pm}
q_{\bm{k}}$.

At this point, it is useful to write down
elementary properties associated with the function 
$\gamma_\pm(x,y)$, all of which are easy to confirm.
\begin{align}
 \gamma_\pm(x,y) & \le \frac{1}{2}(1-r^2)(1+xy) \;\;\text{for} \;
|x|\le 1, |y|\le 1. \label{eq:upperg} \\
\frac{1+xy}{1-r} & \ge \frac{1-xy}{1+r}
\;\;\text{if} \;
\gamma_+(x,y)\ge 0 \;\text{or} \;
\gamma_-(x,y)\ge 0.
\end{align}
Using the latter property, the linearity of the functionals $p$ and $q$, and Eq.~(\ref{eq:equivalence}),
we obtain 
\begin{align}
p_\pm& :=
\sum_{\bm{k}\in \Gamma_\pm} p(\hat{G}_{\bm{k}})
=
\sum_{\bm{k}'\in \Gamma'_\pm} p(\hat{G}_{\bm{k}'})\nonumber \\
& \ge \frac{1-r}{1+r}  \sum_{\bm{k}'\in \Gamma'_\pm} q(\hat{G}_{\bm{k}'})
= \frac{1-r}{1+r} q_\pm . \label{eq:ppm}
\end{align}

Define a linear functional $f_\pm(\hat{K}):= {\rm Tr}[\hat{F}_\pm\hat{K}]/4$ 
with $\hat{F}_\pm:= (\hat{Z}\mp r \hat{1})\otimes (\hat{Z}\pm r
\hat{1})$,
where $\hat{Z}=\ket{0}\bra{0}-\ket{1}\bra{1}$.
For $\hat{G}_{\bm{k}}$ in the form of Eq.~(\ref{eq:matrix}),
we have 
$
 f_\pm(\hat{G}_{\bm{k}})=w_{\bm{k}}\gamma_\pm(x_{\bm{k}},y_{\bm{k}}).
$
For $\alpha>0$, define slightly enlarged regions 
$R_{\pm}^\alpha:= \{(x,y)|\gamma_\pm (x,y)\ge -\alpha (1+xy)\}$.
Then, using Eqs.~(\ref{eq:equivalence}) and (\ref{eq:upperg}), we have
\begin{align}
0\le &\sum_{\bm{k}'\in \Gamma'_\pm} f_\pm(\hat{G}_{\bm{k}'})
=\sum_{\bm{k}\in \Gamma_\pm} f_\pm(\hat{G}_{\bm{k}})
\nonumber \\
\le &- \sum_{\bm{k}\in \Gamma_\pm: (x_{\bm{k}},y_{\bm{k}})\notin R_{\pm}^\alpha}
\alpha p_{\bm{k}}
+
\sum_{\bm{k}\in \Gamma_\pm: (x_{\bm{k}},y_{\bm{k}})\in R_{\pm}^\alpha
}
\frac{1-r^2}{2} p_{\bm{k}}
\nonumber \\
=& -\alpha p_\pm + \frac{1-r^2+2\alpha}{2}
\sum_{\bm{k}\in \Gamma_\pm: (x_{\bm{k}},y_{\bm{k}})\in R_{\pm}^\alpha
} p_{\bm{k}} .
\end{align}
Using Eqs.~(\ref{eq:eta}) and (\ref{eq:ppm}), we have
\begin{equation}
 \sum_{\bm{k}: (x_{\bm{k}},y_{\bm{k}})\in R_+^\alpha\cup R_-^\alpha 
} p_{\bm{k}}\ge 
\frac{2\alpha}{1-r^2+2\alpha}
\frac{1-r}{1+r} Q.
\label{eq:p_lower}
\end{equation}
Let us assume that $\sqrt{Q/(2-Q)}<r<1$ and 
$0<\alpha<[(2-Q)r^2-Q]/2$. Then the region 
$R_+^\alpha\cup R_-^\alpha$ does not include the points
$x=y=\pm \sqrt{Q/(2-Q)}$. From the continuity of 
$\Delta (x,y)$, we have
\begin{equation}
 \Delta_{\min}:= \min_{(x,y)\in R_+^\alpha\cup R_-^\alpha}
\Delta (x,y)>0. \label{eq:minimize}
\end{equation}
Combining Eqs.~(\ref{eq:difference}), (\ref{eq:p_lower}) and
(\ref{eq:minimize}) proves the existence of a nonzero gap,
\begin{eqnarray}
 \bar{E}_{\rm opt}^{\rm sep}&=&
{\cal E}(1-Q)\ge \bar{E}_{\rm opt}^{\rm LOCC} + \Delta_{\rm low} ,
 \label{eq:gap}
\\
\Delta_{\rm low}&:=&
\frac{2\alpha}{1-r^2+2\alpha}
\frac{1-r}{1+r} Q \Delta_{\rm min}>0,
\end{eqnarray}
in the quantum task.

In the above proof for the nonzero gap, the achievability of 
$\bar{E}_{\rm opt}^{\rm sep}$ was given by an example with diagonal 
operators, and the upper bound on $\bar{E}_{\rm opt}^{\rm LOCC}$
was derived by focusing only on diagonal terms of the POVM. This 
allows us to find a task in a purely classical setting that shows 
a similar gap. In doing so, we replace entanglement by privacy,
and LOCC by public communication (PC).
We consider the following task:
(i') Claire privately sends random bits $i$ and $j$ to Alice
and Bob, respectively.
(ii') Alice and Bob try to distinguish the cases with 
$i\neq j$ by announcing an outcome $\bm{k}$ using 
the allowed communication resources specified below.
Let $P(\bm{k}|ij)$ be the probability of $\bm{k}$
conditioned on the bit values $i,j$. Define 
$p_{\bm{k}}^{\rm cl}:=[P(\bm{k}|00)+P(\bm{k}|11)]/2$
and  
$q_{\bm{k}}^{\rm cl}:=[P(\bm{k}|01)+P(\bm{k}|10)]/2$.
If the outcome $\bm{k}$ satisfies $p_{\bm{k}}^{\rm cl}=0$,
the protocol ends.
The discrimination must be efficient enough to satisfy
$\sum_{\bm{k}:p_{\bm{k}}^{\rm cl}=0}q_{\bm{k}}^{\rm cl}\ge Q$.
(iii') Otherwise, Claire announces whether $i=j$ or not.
We ask how much private correlations are left in the case of 
$i=j$. The privacy can be quantified by a function $K(\lambda^{\rm cl})$,
where $\lambda^{\rm cl}$ is  the probability of $i=j=0$ 
conditioned on the publicly announced variables.
Here we adopt $K(\lambda^{\rm cl})={\cal E}(2\sqrt{\lambda^{\rm cl}(1-\lambda^{\rm cl})})$,
using the function ${\cal E}(C)$  used in the quantum case, 
satisfying Eq.~(\ref{eq:Econdition2}). 
The mean residual privacy is then given by
\begin{equation}
\bar{K}:= \sum_{\bm{k}} p_{\bm{k}}^{\rm cl}
K(\lambda_{\bm{k}}^{\rm cl}),
\end{equation}
with $\lambda_{\bm{k}}^{\rm cl}=P(\bm{k}|00)/(2p^{\rm cl}_{\bm{k}})$.

Let $\bar{K}_{\rm opt}^{\rm PC}$ be the supremum over  
the values of $\bar{K}$ achievable via PC at step (ii'),
in which 
Alice and Bob alternately announces $k_1, k_2,\ldots,k_n$.
We can prove
$\bar{K}_{\rm opt}^{\rm PC}\le \bar{E}_{\rm opt}^{\rm LOCC}$
as follows. Suppose that a PC protocol achieves 
$\bar{K}=\bar{K}^*$. 
Then we may construct an LOCC protocol in the quantum 
task by choosing Bob's Kraus operators on his qubit
on the $2s$-th turn to be 
$\hat{B}_{k_{2s}}=\sqrt{\beta_0}\ket{0}\bra{0}_B+ \sqrt{\beta_1}\ket{1}\bra{1}_B$,
with $\beta_j$ determined from Bob's strategy at the 
$2s$-th turn in the PC protocol as 
$\beta_j=P(k_{2s}|j,k_1\cdots k_{2s-1})$.
Alice's operators for her turns are chosen similarly.
Then the final state should be 
$\ket{\psi_{\bm{k}|\pm}}_{AB}=\sqrt{\lambda_{\bm{k}}^{\rm cl}}\ket{00}_{AB}
\pm \sqrt{1-\lambda_{\bm{k}}^{\rm cl}}\ket{11}_{AB}$,
leading to $E(\ket{\psi_{\bm{k}|\pm}}_{AB})=K(\lambda_{\bm{k}}^{\rm cl})$.
It is also easy to see that $p_{\bm{k}|\pm}=p_{\bm{k}}^{\rm cl}$
and $q_{\bm{k}}=q_{\bm{k}}^{\rm cl}$.
Thus the LOCC protocol achieves $\bar{E}=\bar{K}^*$,
and hence $\bar{K}_{\rm opt}^{\rm PC}\le \bar{E}_{\rm opt}^{\rm LOCC}$.

We introduce `separable operations' in the classical setting as
follows. Alice and Bob privately 
submit their bit values $i, j$ to a helping agent, who announces an outcome 
$\bm{k}$ with a probability in a separable form,
$p(\bm{k}|i j)= w_{\bm{k}}[1+(-1)^{i} x_{\bm{k}}][1+(-1)^{j} y_{\bm
k}]$.
It is straightforward to see that the agent 
has no ability to produce a private correlation 
if and only if the probability is written in the separable form. 
Hence the classical separable operation has a clear operational 
meaning as in the case of its quantum counterpart.
Let $\bar{K}_{\rm opt}^{\rm sep}$ be 
the maximum of $\bar{K}$
over the separable operations at  step (ii').
Construct an example of $p(\bm{k}|ij)$ $(\bm{k}=1,2,3,4)$
by choosing the parameters as in Eq.~(\ref{eq:separable-opt}).
Direct calculation shows that $\bar{K}={\cal E}(1-Q)$, implying 
$\bar{K}_{\rm opt}^{\rm sep}\ge {\cal E}(1-Q)$.
Together with the gap in the quantum case, Eq.~(\ref{eq:gap}),
we conclude $\bar{K}_{\rm opt}^{\rm sep}-\bar{K}_{\rm opt}^{\rm PC}
\ge \Delta_{\rm low}>0$, namely,
there is a gap even in a purely classical setting.

\begin{figure}[t]
\includegraphics[keepaspectratio=true,height=35mm]{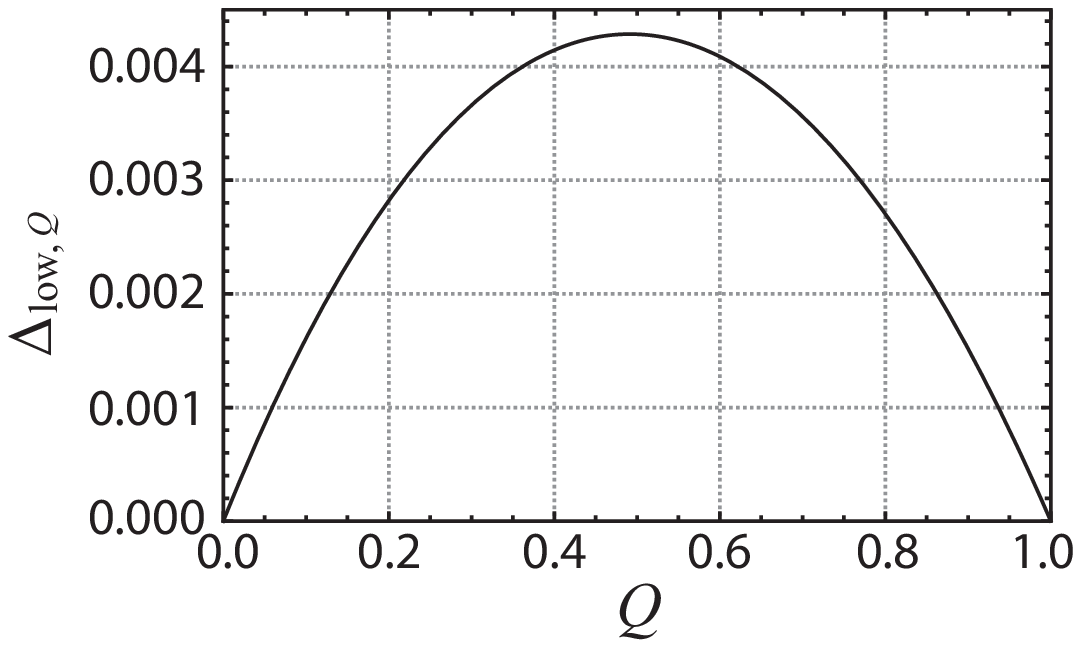}
  \caption{Gap $\Delta_{{\rm low},Q}$ for ${\cal E}={\cal E}_{Q}$.}
  \label{fig:gap}
\end{figure}

Finally, 
assuming ${\cal E}={\cal E}_Q$, we estimate the shared gap 
$\Delta_{\rm low}=\Delta_{{\rm low},Q}$.
In evaluating Eq.~(\ref{eq:minimize}),
from the symmetries of $\Delta (x,y)=\Delta (-x,-y)$ and $\Delta (x,y)=\Delta (y,x)$, 
we can assume $(x,y) \in R' \cap R_{+,\alpha}$ 
with $R' := \{(x,y)| -x \le y \le x\}$. 
Let $(x^*,y^*)$ be thge point satisfying 
 $C(x^*,y^*) = 1-Q$ and $\gamma_+(x^*,y^*) = -\alpha (1+x^* y^*)$
with $x^* < r/(1+\alpha)$. 
Since any $\mu$ with $0<\mu <1/(1-Q)$ satisfies Eq.~(\ref{eq:Econdition2}) for ${\cal E}_Q$, 
we assume $\mu = (1+ x^* y^*)/(1- x^* y^*)$ such that $\Delta(1, 1)=\Delta(x^*, y^*)$.
Then, in the region  $\{(x,y)| C(x,y) \le 1-Q \}$,
we have
\begin{multline}
\frac{\Delta (x, y)-\Delta (x^*, y^*)}{{\cal E}_Q (C(x,y))} = \frac{\mu}{{\cal E}_Q(C(x,y))} \frac{1-x y}{1+xy}-1  \\
= \frac{(1-x^*)(1-y^*)}{1-x^*y^*} \frac{1-xy}{(1-x)(1-y)}-1 \ge 0.
\end{multline}
In addition, a direct calculation shows that $\Delta(x^*, y^*)$ is 
also the minimum in the region $ \{(x,y)| C(x,y) \ge 1-Q \}$.
Thus, Eq.~(\ref{eq:minimize}) becomes $\Delta_{\min} = \Delta(x^*, y^*)$. 
The remaining parameters $r$ and $\alpha$  may be chosen so as to maximize $\Delta_{\rm low}$ in Eq.~(\ref{eq:gap}).
This maximized gap $\Delta_{{\rm low},Q}$ for ${\cal E}={\cal E}_Q$ is shown in Fig.~\ref{fig:gap}.

The gap found in a purely classical setup may 
give us a renewed insight into the origin of the gaps 
that have been discussed since its first discovery \cite{QNWE}.
In quantum tasks under LOCC, each party alternately 
reveals partial information on their local state
through a measurement.
 One may then ascribe the relative inefficiency 
of LOCC to the measurement backaction to the complementary 
observables, leading to disturbances on nonorthogonal states or
degradation of entanglement.
But the current example suggests a much simpler reasoning
without any reference 
to properties of quantum mechanics, that is, it is 
the very act of revealing information on the local state
that makes LOCC inferior to separable operations.

We thank F.~Morikoshi and T.~Yamamoto for valuable discussions.
This research is supported by a MEXT Grant-in-Aid for Scientific Research on Innovative Areas (No. 21102008 and No. 20104003), and by the Japan Society for the Promotion of Science (JSPS) through its Funding Program  for World-Leading Innovative R\&D on Science and Technology (FIRST Program).

\end{document}